\documentstyle[12pt]{article}    
\input epsf  %allows eps files to be inserted easily
\renewcommand{\theequation} {{\arabic{section}}.{\arabic{equation}}}
\begin{document}                    
%%%%%%%%%%%%%%%%%%%%%%%%%%%%%%

%Title: 
\centerline{\Large{\bf  A Hybrid Algebraic/Schr\"odinger Approach to}}
\centerline{\Large{\bf the 2D Franck-Condon Problem}}

%Authors:  
\vspace{.2in}
\centerline{Michael W.N. Ibrahim}

%Author Address:
\vspace{.2in}
\centerline{\em Center for Theoretical Physics,
                Sloane Laboratory,}
\centerline{\em Yale University, 
                                New Haven, Connecticut\ \ 06520--8120}

%Abstract:
\vspace{.5in}
\centerline{\bf Abstract}
We use recent results regarding the geometry of the $U(3)$ bosonic algebraic model to motivate a hybrid algebraic-Schr\"odinger approach to the 2D Franck-Condon problem analogous to 1D approaches.  This approach allows one to analyze bent to linear Franck-Condon transitions of triatomic molecules as well as clarifies the angular momentum dependance of the Franck-Condon intensities.
\vspace{.5in}

%Main Text:

\section{Introduction} \setcounter{equation} {0}

The study of transition amplitudes for molecules with initial and final states given by different electronic configurations  has held great interest.  In the Franck-Condon limit one models the two configurations by similar effective potentials with different geometric parameters.  The transition intensities are then given by the wavefunction overlap of the two potentials (sudden approximation) [\ref{CDL}].  

Recently, the Franck-Condon (FC) problem of polyatomic molecules has been made tractable by the introduction of a hybrid algebraic-Schr\"odinger approach [\ref{II}--\ref{IL}].  In this approach 
 a spectrum generating algebra of $U_1 (2) \times \cdots \times U_k (2)$ (where there are $k$ bonds) is used to obtain 
 wavefunctions [\ref{IOii}]
\begin{equation}
| \psi \rangle = \sum_{i_1,\ldots,i_k} c_{i_1,\ldots,i_k}   |[N_1],i_1 \rangle \otimes \cdots \otimes |[N_k],i_k \rangle,
\end{equation}
  by fitting spectra of a molecule in a particular configuration.
In general for the two different configurations the $i$th bond contribution could change from the representation $[N_i]$ to $[N_i ']$.
The Franck-Condon factors for the polyatomic case are realized in terms of the single bond factors by calculating the matrix element $\langle \psi '|\hat U| \psi \rangle$ [\ref{II},\ref{MDVPII}], where 
\begin{equation}
\hat U =\hat t(\frac{\alpha_1}{\alpha_1 '},\Delta_1) \otimes \cdots
        \otimes \hat t(\frac{\alpha_k}{\alpha_k '},\Delta_k)
\end{equation}
 and the operator $\hat t$  is defined by its matrix elements as determined by Schr\"odinger overlaps:
\begin{equation}
\label{eqn:littlet}
\langle [N'],n'|\, \hat t(\frac{\bar \alpha}{\bar \alpha'},\Delta) \, |[N],n \rangle = 
         \int dx \, \psi^{N'}_{n'}(\bar \alpha'; x)
         \psi^{N}_{n}(\bar \alpha; x- \Delta).
\end{equation}
The parameters $\alpha \over \alpha'$ and $\Delta$ are typically fit
from transition spectra.  Due to the $U(2) \supset O(2)$ chain's
correspondence with a Morse oscillator [\ref{MANY}] one often uses
Morse eigenfunctions with anharmonicity parameter $1 \over N+1$ for
Schr\"odinger wavefunctions.  It is worth noting that the heuristic
approximation in [\ref{II}] showed that this can be equivalently
thought of as simple harmonic oscillator (SHO) overlaps where the
scale parameters acquire a linear $n$ dependent correction with
coefficient $O(1/N)$. That is, one would use the equation for the
overlap of SHO wavefunctions, which has no $N$ dependence, and
substitute a scale $\alpha = \alpha_0 (1 - \xi (n +\frac{1}{2}))$.
$\alpha_0$ was taken as the scale of the SHO best approximating the
Morse and $\xi$ is $O(1/N)$.  In this way one introduces geometric
parameters not contained in the algebra.

Although this approach has been successful it necessitates that the
initial and final configurations not be too dissimilar---in particular
they must have the same normal mode expansion.  This precludes, for
example, a transition between a linear and bent triatomic molecule
(Figure~1).  In the linear configuration the center atom has two
normal modes.  In the bent configuration one of the normal mode
solutions is spurious---corresponding to an overall rotational degree
of freedom.  Because of this, the treatment of vibrational modes in
each configuration necessitates the freezing of a different number of
degrees of freedom, whereas when studying transitions between each
configuration one needs to have the same degrees active.

One may attack this problem by using more complicated algebraic models
which have both geometries built in.  In this way some parameters
which before were artificially inserted through the hybrid method
become natural---i.e. since they are built in the algebra they are
determined by spectra and do not need to be fit with transition data.

However, in these more complicated algebraic models, exact
Schr\"odinger correspondences don't in general exist.  Thus, this
approach necessitates an alternate interpretation of algebraic
parameters as geometric, configuration space quantities.  Recently
such an interpretation has been provided [\ref{ME}] by exploiting the
many approximate correspondences between algebraic and Schr\"odinger
pictures.  We use these results to develop a hybrid approach for the
bent to linear transitions in Figure~1.  This is most easily done by
using a spectrum generating algebra of $U_1 (2) \times U(3) \times U_2
(2)$ (a $U(2)$ for each bond and the $U(3)$ for the two additional
degrees of freedom of the center atom).  The FC transition operator
would then be $\hat U =\hat t_1 \otimes \hat T \otimes \hat t_2$,
where $\hat t_1$, $\hat t_2$ are defined by
equation~\ref{eqn:littlet}.  It is the purpose of this publication to
motivate the definition of $\hat T$ (eqn.~\ref{eqn:FINALE}).

We begin by demonstrating, via a coherent state limit analysis
[\ref{ZFG}], that $U(3)$ is an appropriate algebra to describe the
degrees of freedom of the center atom of the triatomic.  We proceed
with numerical studies to study the implications and discrepancies of
this interpertation.  Finally, we reconcile these discrepancies using
[\ref{ME}] which naturally motivates our definition of $\hat T$.

\section{The Franck-Condon Problem and $U(3)$} \setcounter{equation} {0}

\subsection{Overview of $U(3)$ Statements on 2D Problem}

Th algebraic approach for 2D problems was presented by [\ref{IO}].  One considers symmetric (bosonic) representations of $u(3)$.  There are two chains of interest:
\begin{eqnarray}
\begin{array}{ll}
U(3)\supset U(2) \supset O(2) & {\rm I} \\
U(3)\supset O(3) \supset O(2) & {\rm II}
\end{array}.
\end{eqnarray}
We use the same notation as [\ref{IO}] for the generators except chose
a different $O(3)$ subgroup as explained in [\ref{ME}]. Please note
that the $O(3)$ group is a dynamical symmetry subgroup and does not
have the interpretation of a rotation in configuration space.

The general Hamiltonian of the $U(3)$ model is
\begin{equation}
\label{eqn:U3hamiltonian}
H = \epsilon \, \hat n + \delta  \,\hat n (\hat n +1)
    + \beta \, \hat l^2 - A \, \hat W^2 ,
\end{equation}
 where $\epsilon$, $\delta$, and $A$ are taken as positive or $0$.
 Setting $A=0$ ($\epsilon = \delta = 0$) gives a Hamiltonian with
 dynamical symmetry I (II).  The spectra for each dynamic symmetry may
 be determined from the well known solution to the branching problem
 for a symmetric representation of $u(3)$ labelled by $N$ (the
 eigenvalue of the $u(3)$ Casimir $\hat{n} +\hat{n}_s$) [\ref{IO}].
 The basis corresponding to chain I is labeled by the eigenvalues of
 the $u(2)$ and $o(2)$ Casimirs $n$ and $l$ respectively.  The basis
 corresponding to chain II is labeled by the $o(3)$ Casimir's
 eigenvalues, $\omega(\omega+1)$, and again by $l$.

The spectra of each chain led the authors of [\ref{IO}] to the interpretation of each dynamical symmetry as an azimuthally symmetric potential with minimum at 0 radius (chain I) and at non-zero radius (chain II).

\subsection{$U(3)$ Coherent State Limit}
\label{sec:U3limit}

The interpretation of [\ref{IO}] is reaffirmed by simply studying the classical coherent state limit calculated in [\ref{ME}]. 

Taking the coherent state limit of the Hamiltonian \ref{eqn:U3hamiltonian}, setting all momenta to zero and dropping additive constants one finds the potential in group coordinates [\ref{OriginalGilmore}] to be (up to a meaningless multiplicative factor):
\begin{equation}
\tilde V_{\rm cl}(r)
  = \eta
  \frac{1}{2} r^2
 + \frac{1}{4}\, r^4  
\end{equation}
where
\begin{equation}
\eta = {\epsilon + 2 \delta + \beta - 4A(N-1)  \over (\delta +4 A)(N-1)}.
\end{equation}

It is easy to compute the position of the potential minima 
\begin{equation}
r_{\rm min}=\left\{\matrix{0\, \, &\, \, \eta \geq 0\cr {\sqrt {- \eta}} \, \, &\, \, \eta <0\cr }\right. .
\end{equation}
Thus we find:
\begin{equation}
\tilde V_{\rm cl}(r_{\rm min})=\left\{\matrix{0\, \, &\, \, \eta \geq 0\cr -{1 \over 4}\eta ^2 \, \, &\, \, \eta <0\cr }\right. .
\end{equation}
That is we have a second-order phase transition at $\eta = 0$, or equivalently $4A(N-1) = \epsilon + 2 \delta + \beta $ [\ref{PHASE}].

The exact same analysis may be carried out in projective coordinates [\ref{OriginalGilmore}] revealing:
\begin{equation}
\tilde V_{\rm cl'}(\tilde r_{\rm min})=\left\{\matrix{0\, \, &\, \, \eta \geq 1\cr -{1 \over 4}{(1- \eta)^2 \over 1+\eta'} \, \, &\, \, \eta <1\cr }\right. ,
\end{equation}
where now
\begin{equation}
\eta = {(\epsilon + 2\delta + \beta) \over  4A (N-1)}\, 
\;, \; \; \; \;
\eta' = {\delta \over 4A} \, ,
\end{equation}
and
\begin{equation}
\tilde r_{\rm min}^2=\left\{\matrix{0\, \, &\, \, \eta \geq 1\cr { 1- \eta \over  \eta + 2 \eta' +1} \, \, &\, \, \eta <1\cr }\right. .
\end{equation}
Thus we again find a second-order phase transition at  $4A(N-1) = \epsilon + 2 \delta + \beta $.

In either coordinates the potential minima moves from $r=0$ (corresponding to a linear configuration) to $r \neq 0$ (corresponding to a bent configuration) at the critical point.  We conclude that the algebraic model is rich enough to include both geometries depicted in Figure~1.  Algebraic Hamiltonians having nearly a $U(2)$ dynamical symmetry correspond to the center atom in a linear triatomic, whereas those near the $O(3)$ limit correspond to the center atom in a
bent triatomic

\subsection{$U(3)$ and Schr\"odinger FC Connections}
\label{sec:questions}

The FC factor for the two configurations is easily studied from a
Schr\"odinger perspective.  Assuming that, whatever the actual nature
of the potentials, they may be approximated about their minima as
harmonic we may numerically calculate the FC overlaps.  The results of
such a calculation are displayed in Figure~2.  The parameters
(Table~2) are chosen to be relevant to the bent to linear FC
transition ($\ ^1 B_2 \rightarrow \Sigma_g ^+$) of $CS_2$.  The
frequency of the linear configuration was taken from reference
[\ref{PMSJ}].  The harmonic distance scale, $\frac{m \omega}{\hbar}$,
was deduced by assuming the effective mass was that of the carbon
nucleus.  The radial displacement was deduced from the geometry of the
bent configuration as published in reference [\ref{JMM}] assuming that
the heavy ($S-S$) axis was essentially stationary.  The associated
distance scale of the bent configuration was assumed to be the same as
the linear scale.  Since $CS_2$ is a particularly shallow molecule we
have included an additional plot (Figure~3) to demonstrate the
behavior for a larger radial displacement.

If one models the potential as exactly harmonic the displaced oscillator potential has the idiosyncrasy of a `cusp' at $r=0$.  This is of little concern since the effective potential is dominated by the angular momentum barrier at this point.  Approximate analytic expressions may be obtained for this limit as detailed in Appendix~A.

We wish to emphasize that although these graphs serve as a valid starting point for a more complete analysis of $CS_2$ transition intensities their primary purpose here is heuristic.  A full analysis requires a careful fitting of the bent configuration distance scale.  Additionally, the bending modes considered here are known to couple strongly to symmetric stretching modes [\ref{PMSJ}]---i.e. a full analysis would require coupling additional $U(2)$'s as discussed in the introduction.  

One may consider more realistic bending potentials such as a P\"oschl-Teller:
\begin{equation}
V =  V_0 \left[ {1- \cosh^{-2} \bar \alpha (r-r^*)}\right] \, .
\end{equation}
In this case the cusp at the origin still exists for the displaced
oscillator (non-zero $r^*$) but is tamed due to the long distance
flattening of the potential.  In the limit where the minima is far
from the origin the cusp essentially vanishes.  Figure~3 shows
numerical results for a P\"oschl-Teller model of a bent to linear
transition.  The parameters are chosen such that the the potentials
are approximated to second order by exactly the harmonic plots
included in the same figure.  That is, the harmonic distance scale,
$\alpha^4 = 2 m V_0 \bar \alpha^2 / \hbar^2$ is set to the same value
as the SHO FC factors.  The remaining parameter, taken as $\alpha^2
\over \bar \alpha ^2$, is a unitless measure of well depth.  It was
chosen to be sufficiently small to emphasize differences between the
FC factors of the two potentials.

The previous section's analysis implies that the FC factors for a
bent to linear configuration in the algebraic picture are given by
exactly the inner product of the algebraic wavefunctions for
hamiltonians near the $O(3)$ chain (bent configuration) and on the
$U(2)$ chain (linear configuration).  The overlaps for several such
`bent' hamiltonians are given in Figure~4.  The algebraic `bent'
hamiltonian was taken to be of the form
\begin{equation}
H = (1- \xi ) \hat n - {\xi \over (N-1)} \hat W^2 .
\end{equation}
The parameter $\xi$ was chosen to match the intensity maximum with
that of several harmonic Schr\"odinger calculations (corresponding to
unitless radial displacements of $3$, $5$, and $7$).  The results of
calculations for two significantly different irreps., $[N]$, are shown
to emphasize that the structure is generic and not a function of any
special choice of parameters.

Comparing Figures~2--4 again reaffirms the interpretation of the two
$U(3)$ chains as bent and linear configurations of a 2D problem.
Comparing the SHO and P\"oschl-Teller figures one notes although not
in exact agreement they are very similar given the large differences of the
Schr\"odinger potentials.
Qualitatively the $U(3)$ graphs also appear similar with the possible
exceptions of (1) their (expected) truncation at higher $n$; (2) their
dramatically sharper peaks than the Schr\"odinger FC graphs; (3) their
amplitude's diminished sensitivity to the amount of radial
displacement.  

\subsection{Scale Changes}

The numerical study raises two questions (1) What are the relations
between the algebraic parameters (determining the hamiltonian's
proximity to either chain) and the Schr\"odingers (more geometric)
parameters? (2) Is there an interpretation for the apparent
qualitative differences (the sharper peak) between the algebraic and
Schr\"odinger pictures?

Both these questions can be addressed by considering the results of
[\ref{ME}].  The two relevant results, which we reproduce here,
include the intrinsic distance scale of the harmonic approximation to
a Hamiltonian near the $O(3)$ limit and an expression for the radial
displacement for the same Hamiltonian:

\begin{eqnarray}
\label{eqn:scale}
\alpha^2 \approx  
2 \frac{ \overline{m\omega }_{O(3)} }{ \hbar }
 \left[ 
     1 
          +{\frac{X}{Z}  \left(\frac{l^2}{N^2}-1 \right)
          - \frac{Y}{Z} \left(\frac{l^2}{N^2}  + \frac{1}{2} \right)
           }  \right] .
\end{eqnarray}
\begin{equation}
\label{eqn:displace}
(r^*)^2  
\approx \frac{N \hbar}{\overline{m \omega}_{O(3)} } 
  \left( 1 +{\frac{X}{Z}
              ( \frac{ l^2}{(N \hbar)^2}  -1) - \frac{Y}{Z} } 
   \right),
\end{equation}
where  $X= \epsilon + 2 \delta + \beta $, $Y=\delta  \, (N-1)$, and $Z=4 A \, (N-1)$, and the condition that we are near the $O(3)$ limit implies $\frac{X}{Z}$ is small.  The parameter $\overline{m \omega}$ has the interpretation of the ratio of distance to momenta scales, i.e. $\overline{m \omega}=\bar \alpha (m V_0)^{\frac{1}{2}}$ for a potential $V=V_0 f(\bar \alpha x)$.

Additionally the results of [\ref{ME}] imply that Hamiltonians
from either dynamical symmetry not only correspond to different
geometries as implied by \ref{sec:U3limit} but additionaly to
different intrinsic scales (see Appendix~\ref{app:scales}):
\begin{eqnarray}
\label{eqn:scaledependance}
\zeta_n = \frac{\overline{m \omega}_{U(2)}}{\overline{m \omega}_{O(3)}}
  \approx {2 \log 2 \over 2 - \log 2} 
                \left[ 1 + \frac{4}{\log 2 (2- \log 2)} 
                                     \left\{ \frac{3}{2} \frac{\log N}{N} +
                                            (n \log{ \log 2} + c) \frac{1}{N}
                                     \right\}
                                \right] ,  \\
  c  = \log \log 2  - \log \frac{2^\frac{3}{4}}{2 - \log 2} \; \; \; \;
  \; \; \; \; \;  \; \; \; \; \;   \; \; \; \; \;  \; \; \; \; \;   \; \; \; \; \;  \;  \; \;  \; \; \; \; \; 
\end{eqnarray}
This implies that the algebraic overlap of a $U(2)$ Hamiltonian with
an $O(3)$ one is not simply analogous to the overlap of radially
displaced oscillators, but analagous to the matrix elements of an
operator which radially displaces {\em and} dilatates (much like the
operator matrix elements calculated in [\ref{II}]) changing the
natural scale of the problem.  The degree of dilatation depends on the
proximity of the second hamiltonian to either chain.  For chains near
$U(2)$ the dilatation paramater is essentially $1$.  As one moves
nearer to the $O(3)$ chain the dilatation parameter increases,
approaching the value given by equation \ref{eqn:scaledependance}.
This effect must be accounted for in any algebraic or hybrid approach
to the FC problem.

We saw in the introduction that in the 1D problem the scale parameters
may have been thought as harmonic scale parameters with corrections
linear in the quantum number $n$ of order $1/N$.  The scenario is
similar here---except the quantum number is now $n$ and there are
additional corrections of the larger order $\log N / N$.

\subsection{Schr\"odinger and Algebraic Parameter Relations}

Equations \ref{eqn:scale} and \ref{eqn:displace} for the harmonic dilatation and radial displacement establish the needed connection between algebraic and geometrical parameters---at least in the regime where we are near the $O(3)$ limit.  These quantities can be easily related to experimental data.

Experimentally one can find the lower energy level spacing $\Delta E_{\rm exp}$, the reduced mass of the particle $m_{\rm exp}$, and from rotational spectra the distance of displacement $r_{\rm exp}$.  In terms of these quantities one may compute the unitless distance using the harmonic oscillator dilatation%
\begin{equation}
\label{eqn:expSHOdilatation}
\alpha_{\rm exp}^2 = {m_{\rm exp} \Delta E_{\rm exp} \over \hbar ^2}.
\end{equation}
Equating $ \alpha_{\rm exp}^2 r_{\rm exp}^2$ with the previous expressions ($\alpha^2 {r^{*}}^2$ given by \ref{eqn:scale} and \ref{eqn:displace}) one finds
\begin{equation}
2N \left[  1 + {\rm corrections} \right] = {m_{\rm exp} \Delta E_{\rm exp} \over \hbar^2} r_{\rm exp}^2.
\end{equation}
This could be very valuable when fitting spectra.  Since the $O(3)$ chain represents the `maximum' radial displacement [\ref{ME}] this expression gives a lower bound for $N$.  One may begin fitting data for the $N$  which satisfies this equation with the corrections set equal to $0$.  If the $O(3)$ chain spectra with this $N$ doesn't fit the experimental data then one can try higher $N$ and move off the $O(3)$ chain a corresponding amount so that the equation with the corrections is still satisfied.

\section{2-D Franck-Condon Problem:  The Prescription} \setcounter{equation} {0}

\subsection{Adding Dilatations}
\label{sec:dilatations}
We are now ready to develop a prescription for calculating the FC
factors for a $U(3)$ algebraic model.  We begin by considering the FC
problem of two configurations of a molecule both described by the
$U(2)$ chain---i.e. two linear triatomics.  Following the 1D procedure we propose a hybrid approach
based upon calculating the matrix elements of the operator $\hat T$
defined in terms of 2D SHO Schr\"odinger overlaps:
\begin{equation}
\langle [N'], n', l| \hat T |[N], n, l \rangle = T_{n,n',l}(\frac{\alpha}{\alpha'})
\end{equation}
The appropriate overlaps are calculated in Appendix~\ref{app:dilatations} (equations~\ref{eqn:Tmatrix} and \ref{eqn:Tprematrix}).  $\hat T$ by construction does not connect subspaces of different $\l$.

\subsection{Final Procedure}

We have yet to add any dependance on the representation label $N$.  One could add such dependance by hand, appealing to an analogy with the 1D case.  However, when one has a bent to linear transition such an appeal is largely unnecessary due to equation~\ref{eqn:scaledependance}.

For simplicity suppose we have a molecule with a bent configuration whose spectra is fit by a hamiltonian of the $O(3)$ chain in representation $N$ and a linear configuration whose spectra is fit by the $U(2)$ dynamical symmetry with $\delta = \beta = 0$ (in this scenario the second representation label $N '$ can be arbitrary since the hamiltonian will generate identical low lying (experimentally measurable) spectra regardless of $N '$).  Further, suppose the induced harmonic dilatations $\alpha$ and $\alpha '$ (given by equation~\ref{eqn:expSHOdilatation}) can be calculated.  In this instance we know that expanding the $O(3)$ basis in terms of the $U(2)$ basis is equivalent to expanding in terms of a SHO with harmonic dilatation $\alpha_{U(2)}^2 = \frac{1}{2} \zeta_n \alpha^2$ (where $\zeta_n$ is determined by~\ref{eqn:scaledependance}).  Thus the FC transitions should be the matrix elements:
\begin{equation}
\langle [N'], n', l| \hat T |[N], \omega, l \rangle, 
\end{equation}
where $\hat T$ is defined by
\begin{equation}
\langle [N'], n', l| \hat T |[N], n, l \rangle = 
  T_{n,n',l}\left( \frac{\alpha}{\alpha'} \sqrt{\frac{1}{2} \zeta_n} \right). 
\end{equation}
Expanding in large $N$ we see $\sqrt{\frac{1}{2} \zeta_n}= a + b n $ from equation~\ref{eqn:scaledependance} where $b$ is $O(\frac{1}{N})$ and $a$ has constant contributions and $\log N \over N$ contributions.  With the exception that $a \neq 1 $ this is exactly the correction one has in the 1D case.  That is, we have just let $\alpha \rightarrow \alpha (a + b n)$.  

In this extreme case the FC factors can be calculated with no extra fitting parameters.  However, our calculations depended on expansions in large $N$ [\ref{ME}] and thus we would expect them to be inaccurate for larger $n$.  This can be compensated for phenomelogically by allowing $b$ to be fit to compensate for ignored terms.  

Although the above situation is only for a limiting case, these results (along with insight from the 1D analysis) will imply exactly what will happen in other more realistic cases.

\begin{itemize}
\item {\bf Bent configuration near the $O(3)$ basis:}  One can in principle calculate the scale dependance near $O(3)$: $\overline {m \omega}|_{{\rm Off} O(3)} \approx \overline{m \omega}|_{O(3)} ( 1 + \nu \frac{X}{Z} + \mu \frac{Y}{Z})$ where $\nu$ and $\mu$ would have to be determined by repeating the calculation of [\ref{ME}] to second order.  In such a calculation, it is clear that the result will go like  $\alpha \rightarrow \alpha (\tilde a + \tilde b n + \tilde d l^2)$, where $\tilde a = a (1+ \gamma)$ and $\gamma$ is $O(\frac{X}{Z},\frac{Y}{Z})$; $\tilde b = b (1+ \gamma ' )$ and  $\gamma '$ is $O(\frac{X}{Z},\frac{Y}{Z})$; the new parameter $\tilde d$, introducing $l$ dependence, is $O(\frac{X}{Z N^2},\frac{Y}{Z N^2})$; and we have ignored a $n$-$l$ cross term which is of a significantly smaller order.  Although these coefficients could be calculated in principle, one must phenomenologically fit $\tilde b$ and $\tilde d$ in order to describe higher states anyways.  Since the $l$ dependance is of lower order it seems likely that $\tilde d$ may not even be calculable with todays data.

\item {\bf Linear configuration in $U(2)$ basis but $\delta \neq 0$:}  In this instance the $u(2)$ requantization was approximate and inadequate for higher eigenstates.  However, comparing the spectra to a Dunham expansion one sees that $\frac{\delta}{\epsilon}$ plays the role of an anharmonicity parameter (recall equation~\ref{eqn:U3hamiltonian}).  Thus, from the 1D case one concludes that $\alpha ' \rightarrow \alpha ' (1 + b' (n+1))$ where $b'$ must be fit but should be of the order of  $\frac{\delta}{\epsilon}$.

\item {\bf Either configuration far from both chains:}
Although our scheme has not allowed us to do explicit calculations in this regime, in terms of phenomologically fit parameters the result should still be clear.  One should replace the induced harmonic dilatation by $\alpha \rightarrow \alpha ( \tilde a +  \tilde b (n+1) + \tilde d l^2)$.  In this scenario $\tilde a$ should be between $1$ and $a$ depending on the proximity to either chain, $ \tilde b$ has contributions due to both  $\delta$ anharmonicities  and the $O(3)$ chain, and one again expects that $\tilde d$ is of significantly smaller order.  In this regime all parameters must be fit to data.  
\end{itemize}

Let us recapitulate.  The scheme we propose involves fitting the spectra of each configuration of the molecule in a $U(3)$ model to obtain wavefunctions $|[N], \psi_E \rangle$, $|[N'], \psi_{E'} ' \rangle$.  The FC factors are then 
\begin{equation}
\langle [N'], \psi_{E'} ' | \hat T | [N], \psi_{E} \rangle
\end{equation}
where we define $\hat T$ in terms of the $U(2)$ basis:
\begin{equation}
\label{eqn:FINALE}
\langle [N'], n', l'| \hat T |[N], n, l \rangle = \delta_{l,l'} \,
 T_{n,n',l}\left({\alpha (\tilde a + \tilde b (n+1) + \tilde d l^2) \over
                   \alpha' (\tilde a' + \tilde b' (n'+1) + \tilde d' l^2)} \right).
\end{equation}
In general the parameters must be fit, but whenever any of the limiting cases appear the appropriate theoretical values may be substituted. For example, if the linear configuration is highly harmonic one has $\tilde a' = 1$ and $\tilde b' = \tilde d' = 0$.   Further, since we expect the $l^2$ dependance to be nearly negligible and we have a $l = l'$ selection rule, we may expand and combine $\tilde d$ and $\tilde d'$ into one parameter. 

Note that this scheme may also be used for bent to bent transitions.  This would be more useful than the 1D procedure [\ref{II},\ref{MDVPII}] if one was interested in the $l$ dependance of the transitions for instance.

\section{Conclusions} \setcounter{equation} {0}
Using the result of [\ref{ME}] involving the geometry and scales of the dynamical symmetries of $U(3)$ we have explained the difference one sees when considering FC overlaps as described by the $U(3)$ algebra and FC overlaps as described by radially displaced Schr\"odinger oscillators.  Additionally, the analysis has given us the practical result of a minimum value of $N$ as a function of simple experimentally measured quantities.

We observed that this scale dependency leads very naturally to a description of two dimensional FC factors.  To compute the FC factors one expands the algebraic wavefunction in the $U(2)$ basis and formally replaces the basis element's overlap by the 2D SHO Schr\"odinger overlap.  Corrections for $N$, anharmonicities, and even $l$ dependance are made by letting the SHO dilatation constant have $n$ and $l^2$ dependent contributions.  The resulting formulas are analogous to the 1D results obtained by expanding in anharmonicities.

\section{Acknowledgements} \setcounter{equation} {0}
This work was performed in part under the U.S. Department of Energy, Contract No.~DE-FG02-91ER40608.  I extend my deepest gratitude to my advisor, Prof.~Franco Iachello, for introducing me to the FC problem, his suggestions and feedback on this work, and his useful comments on earlier versions of this manuscript.  I am indebted to Thomas M\"uller and Prof. Patrick Vaccaro for useful discussions regarding the status of experimental results on bent to linear FC transitions.  Finally, I would like to thank the Department of Energy's Institute for Nuclear Theory at the University of Washington for its hospitality during the completion of this work.

%%%
%%% Appendices
%%%

\

\

\noindent
{\Huge {\bf Appendices}}

\renewcommand{\theequation} {{\Alph{section}}.{\arabic{equation}}}
\appendix
\section{Approximate Analytic Expressions for 2D FC overlaps} \setcounter{equation} {0}
\label{app:analytic}

\subsection{Linear Configuration}

In the linear configuration in the harmonic limit one has, after separating out the azimuthal portion of the wavefunction, the equation:
\begin{equation}
{1 \over r}{d \over dr}\left({r{d \over dr}v(r)}\right)-{l^2  \over r^2 }v(r)\, +\left({\tilde E-\lambda ^2 r^2 }\right)v(r)=0,
\end{equation}
where $\lambda$ is related to the frequency of the oscillator by $\lambda = {m \omega \over \hbar}$ and $\tilde E = {2m \over \hbar^2}E$.  Subject to the boundary conditions ${{v(r)}|}_{r=\infty }=0$, ${{r v(r)}|}_{r=0 }=0$ the solution is well known to be [\ref{F}]:
\begin{equation}
\label{eqn:SHOwavefunctiions}
v_{n_r,l}(r) = \sqrt{ 2 \; n_r ! \, \lambda^{|l|+1}\over (|l|+n_r)! } \; r^{|l|}e^{-{\lambda \over 2}r^2} L_{n_r}^{|l|}(\lambda r^2),
\end{equation}
with eigenvalue
\begin{equation}
\tilde E = 2 \lambda (|l| +1 +2 n_r).
\end{equation}

\subsection{Bent Configuration}
For a system w/ equilibrium located at some $r=r_0$ we again assume that for the lowest states the system is well approximated by a harmonic oscillator.  Our wave equation is thus:
\begin{equation}
{1 \over r}{d \over dr}\left({r{d \over dr}v(r)}\right)-{l^2  \over r^2 }v(r)\, +\left({\tilde E'-{\lambda '} ^2 (r-r_0)^2 }\right)v(r)=0 .
\end{equation}
Notice, that this mild transformation will greatly change the form of the solutions (as opposed to the 1D case) since  the laplacian is not invariant under radial displacements.  Note that towards the origin the potential is `heightwise truncated' (in full 2-space the potential is not differentiable at the origin).  Thus, to harmonically approximate the potential we must assure that we are sufficiently far from the origin, i.e. loosely ${\lambda '} r_0^2$ must be large.

To obtain approximate solutions to this bent configuration equation we make the substitution $v(r) = {u(r) \over \sqrt r}$.  
For $l \neq 0$ we may expand the effective potential about its minima 
\begin{equation}
r^* = r_0 \left( 1 +   \epsilon - 3 \epsilon^2 +O(\epsilon^3) 
                \right)
\end{equation}
where $\epsilon = \frac{1}{{\lambda '}^2}(l^2-\frac{1}{4}) \frac{1}{r_0^4}$ and
our large $r_0$ condition is refined to
\begin{equation}
\label{eqn:r0condition}
27 r_0^4 >  \frac{256}{{\lambda '}^2}(l^2-\frac{1}{4}).
\end{equation}
Moving the left boundary condition from $r=0$ to $r=-\infty$ only introduces $O(\epsilon)$ corrections  so we may obtain a 1D SHO equation:
\begin{equation}
-{d^2 \over dr^2}u(r)+\left( \Delta + {\bar \lambda}^2 (r- r^*)^2 \right)u(r)=\tilde E'u(r).
\end{equation}
where $\Delta= r_0^2 {\lambda '}^2 (\epsilon - {\epsilon}^2+O({\epsilon}^3))$ and ${\bar \lambda}^2 = {\lambda '}^2 (1 +3 \epsilon -12 {\epsilon}^2 +O({\epsilon}^3))$. Note that the $l$ dependence of the solution is implicit to the behavior of $\bar \lambda$.  We find that the actual wavefunction for the bent configuration is:
\begin{equation}
v_{n_r,l}(r) \approx  {\bar \lambda^{\frac{1}{4}} \over
         (\sqrt{\pi} 2^{n_r} n_r!) ^{\frac{1}{2}}} \; {e^{- \frac{1}{2}{\bar \lambda} r^2} \over \sqrt r} H_{n_r} (\sqrt{\bar \lambda} (r-  r^*)).
\end{equation}
Although this solution is singular at the origin this is of little concern since the metric contains a factor of $r$ (the real objects of interest are $\sqrt r v(r)$, of which we have a good approximation).  

\subsection{Franck-Condon Factors}
We wish to calculate the overlap of wavefunctions from a linear and bent configuration.  Given the $SO(2)$ symmetry of both configurations the angular part of the wavefunctions simply gives the $\delta_{l,l'}$ selection rule.  Hence, the integral we wish to evaluate is
\begin{equation}
I_{n_r, n_r'} =M_{n_r',l}N_{n_r,l} \; \int_0^{\infty} dr \; r^{|l|+\frac{1}{2} } e^{-{\lambda+{\bar \lambda} \over 2}r^2} L_{n_r}^{|l|}(\lambda r^2)    H_{n_r'} (\sqrt{\bar \lambda} (r-  r^*)),
\end{equation}
where we introduced the short hand $M$ and $N$ for the normalizations.

We remove all dimensionful quantities by making the change of variable $u = {\sqrt{\lambda}} r $.  Additionally we decide to work in the (realistic) domain where $\lambda \approx \lambda'$ by setting ${\lambda' \over \lambda} = 1 + \delta$ where $\delta$ is small.  Using expressions  from [\ref{II}] for the expansion
\begin{equation}
H_s({\alpha ({x-x_0})})=\sum_{0\leq n\leq s}t_{n{\rm ,}s}({\alpha {\rm ,}-x_0})H_n(x),
\end{equation}
assuming $\delta$ and $\epsilon$ are of the same order, and ignoring quadratic terms:
\begin{equation}
I_{n_r, n_r'} \approx { M_{n_r',l}N_{n_r,l}\over \lambda^{\frac{|l|}{2} + \frac{3}{4}}} \left({-2{(l^2 - \frac{1}{4})^{\frac{1}{4}} \over \epsilon^{\frac{1}{4}}}}\right)^{n_r'} \; \int_0^{\infty} du \; u^{|l|+\frac{1}{2} } e^{- u^2} L_{n_r}^{|l|}(u^2).
\end{equation}
Not surprisingly the dominant term is independent of $\delta$ (reflecting the fact that FC integrals tend to be more sensitive to changes in displacement than to dilatation).  Evaluating the integral [\ref{GR}] and simplifying:
\begin{equation}
I_{n_r, n_r'} \approx  
        \left({2^{n_r'} \over 2 \sqrt{\pi} \, (|l|+n_r)!\, n_r'! \, n_r!}\right)^{\frac{1}{2}} 
                \left(-r_0 \sqrt{\lambda '} \right)^{n_r'} \; 
                \left(\frac{|l|}{2}+ \frac{1}{4}\right)_{n_r} \;
                \Gamma\left(\frac{|l|}{2} + \frac{3}{4}\right).
\end{equation}

\section{Schr\"odinger and Algebraic Quantum Numbers  } \setcounter{equation} {0}
\label{app:quantumnumbers}
In the preceding sections for each $l = 0, \pm 1, \pm 2 \ldots$ we have that $n_r=0, 1, 2,\ldots$ with the energy relation $E \propto |l| + 2 n_r$.  For the $u(2)$ chain of the $u(3)$ model with hamiltonian $H= E_0 + k \hat n$ (the algebraic model most similar to the 2D SHO) we have as a spectra $E \propto n$.  The branching rules imply that for each $l = 0, \pm 1, \pm 2 \ldots$ we have $n=|l|, |l|+2, |l|+4,\ldots$.  Note that both the expressions for the spectra and the rules for the ranges of the quantum numbers agree if we make the identification:
\begin{eqnarray}
n_r = {n-|l| \over 2} .
\end{eqnarray}
Thus the wavefunction with lowest energy within a subspace of given $l$ is labelled by $n_r=0$ in the Schr\"odinger picture and $n=|l|$ in the algebraic picture.

\section{Scale Changes  } \setcounter{equation} {0}
\label{app:scales}

Our equation \ref{eqn:scaledependance} differs from the results of
[\ref{ME}] by the substitution $|l| \rightarrow n$.  This difference
occurs because [\ref{ME}] did not compute overlaps for higher energy
wavefunctions since its approximation was most valid in the low
energy limit.  However, their requantization technique works exactly
for all levels of the $U(2)$ chain when $\delta = 0$.  We therefore may
expect these higher energy wavefunctions to be more reliable and
repeat their procedure to calculate the scale change for all levels in
this larger domain.  Proceeding to do so one finds that the linear
$|l|$ dependance of equation~\ref{eqn:scaledependance} was an artifact
of working in the ground state of each $|l|$ subspace and the true
correspondence has linear contributions appearing in $n$.

\section{Schr\"odinger FC Factors for dilatated $r^*=0$ Harmonic Potentials} \setcounter{equation} {0}
\label{app:dilatations}

Using the solution to the radial part of the Schr\"odinger equation for the
2D SHO stated in Appendix~\ref{app:analytic} the FC overlaps for a dilatated SHO become:
\begin{equation}
2 \sqrt{  n_r ! n_r' !\, (\lambda \lambda')^{|l|+1} \over (|l|+n_r)! (|l|+n_r')!} \int_0^{\infty} r \, dr \, r^{2|l|} e^{-{\lambda +\lambda' \over 2 }r^2}  L_{n_r}^{|l|}(\lambda r^2) L_{n_r'}^{|l|}(\lambda' r^2) .
\end{equation}
The integral may be evaluated in terms of a hypergeometric function [\ref{GR}] and simplified to:
\begin{eqnarray}
\label{eqn:Tprematrix}
\lefteqn{I_{n_r,n_r',l}^{\rm SHO}(\lambda,\lambda') = \sqrt{(\lambda \lambda')^{|l|+1}\, n_r!\, n_r'! \over (n_r+|l|)! \, (n_r'+|l|)!}(-)^{n_r} 
     \left({2 \over \lambda +\lambda'}\right)^{|l|+1} } \, \, \, \,
\, \, \, \,\, \, \, \,\, \, \, \,\, \, \, \, \\* \nonumber
\, \, \, \, & \times &
         \; \sum_{m=0}^{{\rm min}(n_r,n_r')} {(-)^m (n_r+n_r'+|l|-m)! \over
                                        m! \, (n_r-m)! \, (n_r'-m)!}
                \left( {\lambda-\lambda ' \over \lambda + \lambda '} \right)^
                   {n_r+n_r'-2m},
\end{eqnarray}
where  we rearranged the sum to make the formula slightly more amenable to computer implementation.  With some minor algebra and judicious manipulation of Pochhammers one sees that the expression has the correct $\lambda' \rightarrow \lambda$ limit of $\delta_{n_r,n_r'}$.

Given the relationships of Appendix~\ref{app:quantumnumbers} and that $\lambda = \alpha^2$ the result needed for Section~\ref{sec:dilatations} is:
\begin{equation}
\label{eqn:Tmatrix}
T_{n,n',l}(\frac{\alpha}{\alpha'}) = I_{n_r=\frac{n-|l|}{2},n_r'=\frac{n'-|l|}{2},l}^{\rm SHO}(\alpha^2,{\alpha '}^2)
\end{equation} 
%

%
%
%References and Notes:
%
%

%Figure Captions:
\newpage

\begin{figure}[hp]
\centerline{\epsfbox{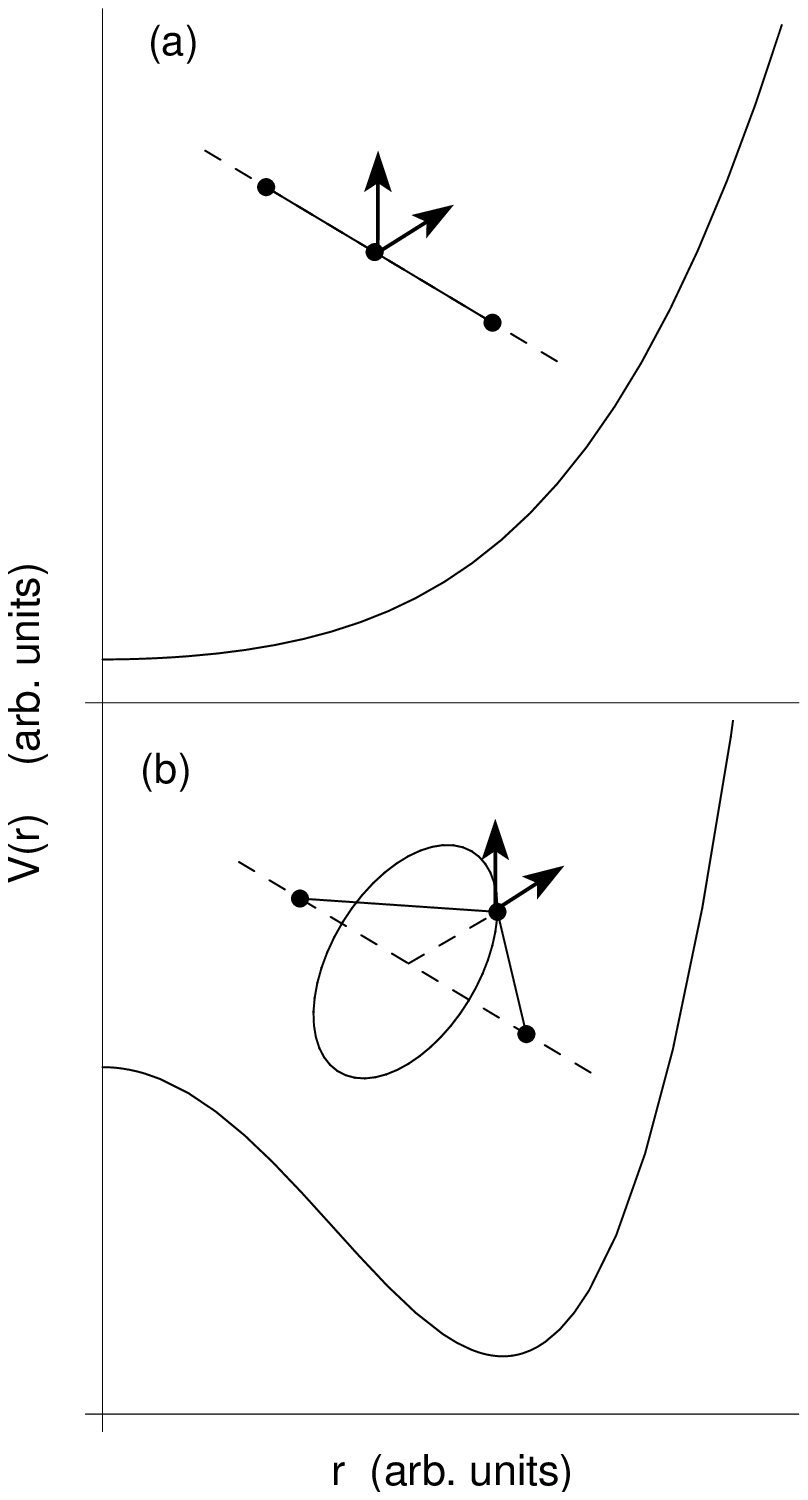}}
\caption {The geometry and radial potentials (of the central atom) of a triatomic model.  In the linear configuration (a) the potential has a true minima at $r=0$ (the Hessian is positive definite) and there is a normal mode for each of the two degrees of freedom.  In the bent configuration (b) the potential has a continuous manifold of minima along a circle surrounding the origin (the Hessian has one positive and one $0$ eigenvalue).  There is only one non-spurious normal mode (radial vibrations or bending modes).  The second mode is lost to the overall rotational degree of freedom.}
\end{figure}

\begin{figure}[hp]
\centerline{\epsfbox{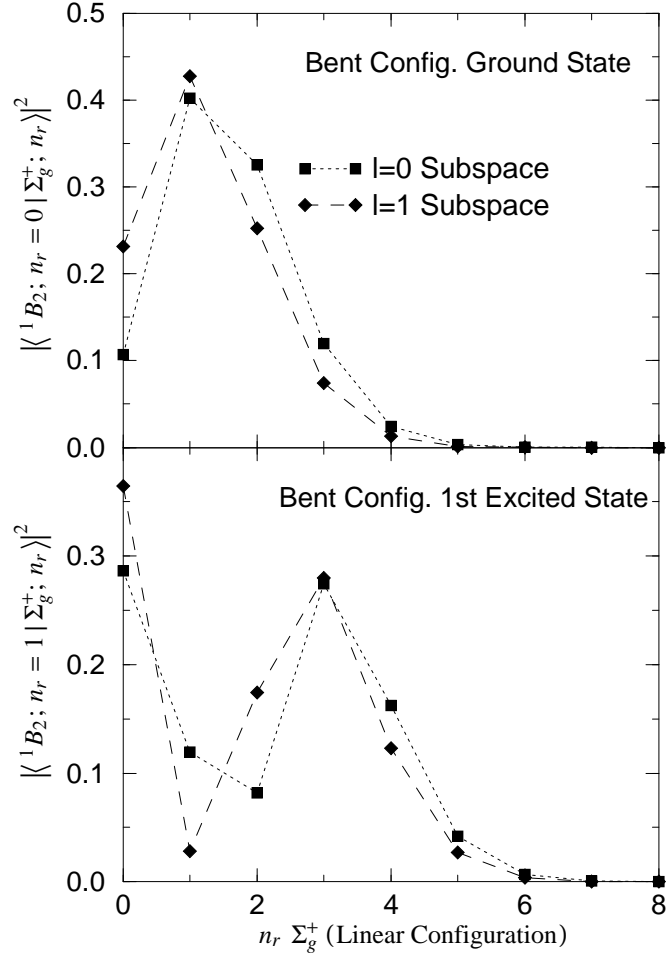}}
\caption {The FC factors from a bent to linear configuration calculated numerically assuming the potential is harmonic about the minima.  The parameters are chosen for the $\ ^1 B_2 \rightarrow \Sigma^+_g $ transition of $CS_2$ (see Table~2).  The factors are plotted for a fixed bent configuration ($\ ^1 B_2$) state as a function of the radial quantum number of the linear configuration ($\Sigma^+_g $) for two subspaces of constant $l$. } 
\end{figure}

\begin{figure}[hp]
\centerline{\epsfbox{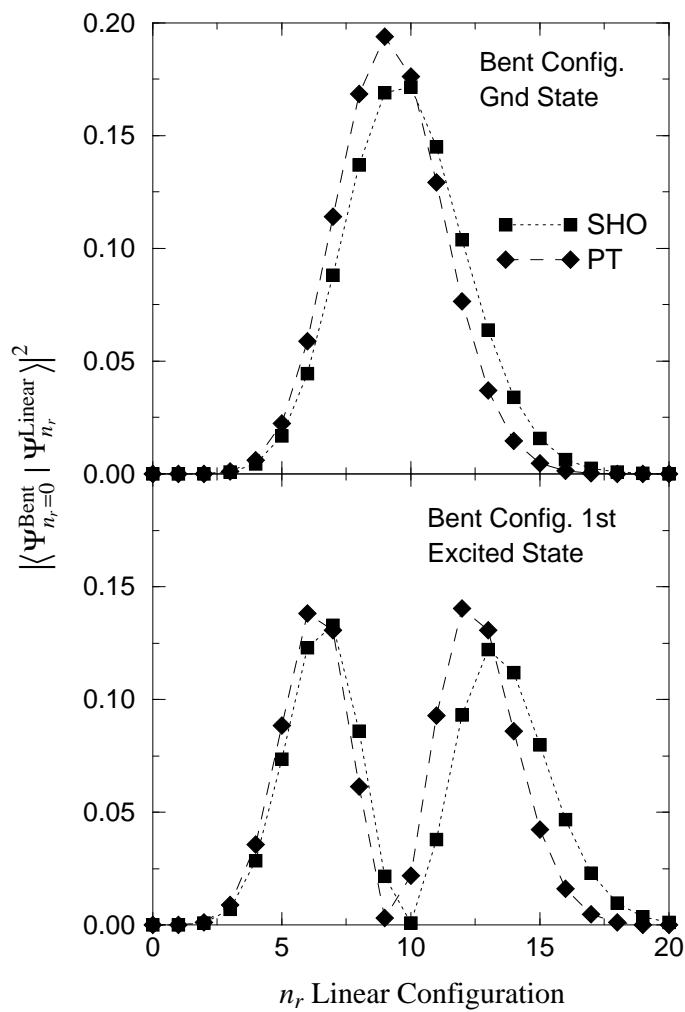}}
\caption {The FC factors from a bent to linear configuration calculated for a fictitious molecule having similar harmonic distance scales as $CS_2$ but a much larger bend (see Table~2). The factors are plotted for a potential of both harmonic (SHO) and P\"oschl-Teller (PT) type.  The depth of the PT potential was chosen so that the graph shows states up to nearly the disassociation energy.}
\end{figure}

\begin{figure}[hp]
\centerline{\epsfbox{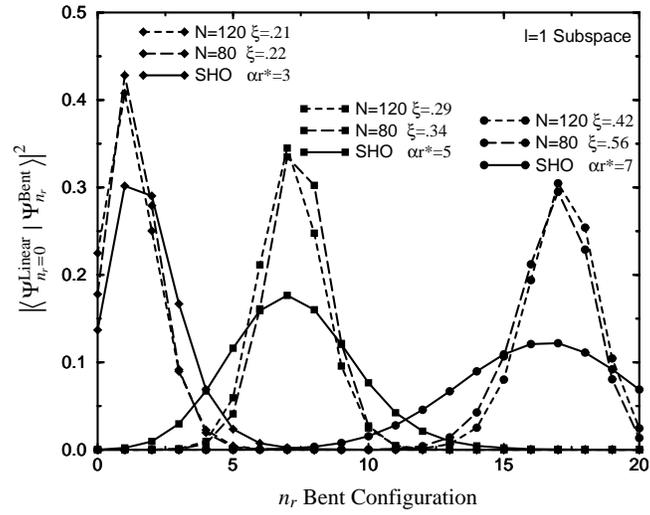}}
\caption {The FC factors from a linear to bent configuration calculated for three different radial displacements assuming the transition corresponds to a (1) $U(3)$ change of basis within the irrep. $[N] = 120$; (2) $U(3)$ change of basis within the irrep. $[N]=80$; (3) harmonic Schr\"odinger potential.}
\end{figure}

%Tables:

\newpage

\begin{table}[hp]
\centerline{ 
\begin{tabular}{|llccc|}   \hline \hline
	 & & SHO (Fig.~2)			& SHO (Fig.~3) & PT (Fig.~3) \\ \cline{3-5}
$\alpha^2 $& \multicolumn{1}{l|}{(${\rm cm}^{-2})$}	& $1.418 \times 10^{18}	$	& $1.418 \times 10^{18}$ & $1.418 \times 10^{18}$ \\
$r_{\rm bent}^*$& \multicolumn{1}{l|}{(${\rm cm})$}	& $ 2.282\times 10^{-9}$	&$ 5.459 \times 10^{-9}$ &$ 5.459 \times 10^{-9}$ \\  
$ \alpha^2 \over \bar \alpha^2 $ & \multicolumn{1}{l|}{ (unitless)} & -- & -- &312. 
\\ \hline \hline
\end{tabular} }
\caption {Parameters for Figures~2 and 3.  The parameters of Figure~2 were chosen to reflect the $\ ^1 B_2 \rightarrow \Sigma^+_g$ transitions of $CS_2$.  The parameters of Figure~3 correspond to a similar molecule with a greater bend.}
\end{table}

\end{document}